# A Dual–Beam Irradiation Facility for a Novel Hybrid Cancer Therapy

Svilen Petrov Sabchevski • Toshitaka Idehara • Shintaro Ishiyama •
Norio Miyoshi •Toshiaki Tatsukawa

*Abstract*: In this paper we present the main ideas and discuss both the feasibility and the conceptual design of a novel hybrid technique and equipment for an experimental cancer therapy based on the simultaneous and/or sequential application of two beams, namely a beam of neutrons and a CW (continuous wave) or intermittent sub-terahertz wave beam produced by a gyrotron for treatment of cancerous tumors. The main simulation tools for the development of the computer aided design (CAD) of the prospective experimental facility for clinical trials and study of such new medical technology are briefly reviewed. Some tasks for a further continuation of this feasibility analysis are formulated as well.

## 1. Introduction

Nowadays, the modern oncologic therapy uses various techniques for cancer treatment such as surgery, radiation and chemotherapy. Among them are several therapeutic techniques for neutron radiotherapy, namely: (i) fast neutron radiotherapy; (ii) neutron capture therapy (NCT), most notably BNCT (boron NCT) [1, 2]; and (iii) a combination of BNCT and a fast neutron radiotherapy [3]. Several other combinations are also known, such as, for example, BNCT supplemented by an external X-ray beam [4]. Both the advantages and the problems associated with these techniques are well-known and the current state-of-the-art has been reviewed recently in [5], where the main challenges and opportunities are discussed. Although both, reactor based and accelerator based, neutron sources [6] are used for production of epithermal (energy ranging from 1 eV to 10 keV) and thermal (energy ca 0.025 eV) neutron beams the latter are considered as the most appealing alternative since they can be integrated more easily and conveniently within the necessary medical infrastructure. Therefore, it is believed that such neutron sources are more preferable for a clinical implementation of the BNCT as a routine modality. Recently developed compact neutron sources (mini neutron tube) at Berkeley Laboratory (see for example [7]) have many advantageous features (small weight and dimensions; portable and highly scalable; safe deuterium-deuterium reaction etc.) and probably will also find application in the

[1,2] S. P. Sabchevski ( ✉ ) • [1]T. Idehara • [3]S. Ishiyama • [4]N. Miyoshi • [5]T. Tatsukawa

[1]Research Center for Development of Far-Infrared Region, University of Fukui, Fukui 910-8507, Japan
 e-mail: idehara@fir.u-fukui.ac.jp

[2]Institute of Electronics of the Bulgarian Academy of Sciences, Association EURATOM-INRNE, Sofia 1784, Bulgaria
 e-mail: sabch@ie.bas.bg

[3]Quantum Beam Science Directorate, Japan Atomic Energy Research Institute, 2-4 Shirakata-Shirane, Tokai, Ibaraki 319-1195, Japan
 e-mail: ishiyama.shintaro@jaea.go.jp

[4]Faculty of Medicine, University of Fukui, Matsuoka-cho, Yoshida-gun, Fukui-ken 910-1193, Japan

[5]Honorary Professor, Kagawa University, Miki-cho, Kida-gun Kagawa-ken 761-0793, Japan

medical technologies. Similar miniature sources even allow the neutron producing target to be put at the end of a tiny catheter–like drift tube [8].

Another promising therapy is based on the irradiation by electromagnetic waves ranging from RF (radio frequency) to sub–terahertz frequencies (sub–millimeter waves). It is based on the physical and physiological action of the electromagnetic fields on the biological material, e.g. living tissue [9–11] through different mechanisms [12–14]. While the radiation in the optical region (visible light) and the infrared region interacts with the electronic levels and the molecular levels, respectively, the sub–millimeter waves interact with the molecular rotations. The specific features of the interaction of terahertz (sub-millimeter) waves with biological systems have been reviewed recently in [15–17]. In this spectral range, the energy of the radiation absorbed in the living cells is converted to a great deal into thermal energy. This process (heating) is the basis of two of the most widely used techniques namely the hyperthermia induced by microwave diathermy (in the range of 40–42 °C) and ablation (at temperatures higher than 50 °C) for treatment of malignant tumors. It should be noticed that the sub–millimeter wave radiation is non-ionizing, and, moreover the energy of a quantum ($h\nu=1.17\times10^{-3}$ eV at $\lambda=1$ mm) is even lower than the energy of the thermal motion at room temperature ($kT=2.53\times10^{-2}$ eV). One of the most characteristic features of the millimeter wave radiation however is its strong absorption by water and aqueous solutions (recall that ca 70 % of the human body is water). For example, a water layer of only 1 mm attenuates the wave with a length of $\lambda=2$ mm $10^4$ times and with a length of $\lambda=8$ mm 100 times. Such strong absorption is due to the fact that the frequency of the rotational motion of water molecules is in the millimeter and sub-millimeter wave length region. Since the cancerous tissue usually has an increased water content the permittivity of the tumors is 10–30 % greater. Additionally, its conductivity is also higher than the surrounding healthy tissue. Thus the conditions for energy absorption in malignant cells and in the normal one is different, which allows a selective treatment of the former. In them, an increased temperature induces cellular death via coagulation necrosis and, eventually, to a tumor ablation while leaving intact the viable healthy cells.

It should be noted, however, that the thermal effects are not the only consequences resulting from the irradiation of biological tissue by electromagnetic waves in the sub-terahertz frequency range [18–21]. Despite the fact that there is an ongoing debate concerning the underlying mechanisms of the so called non–thermal (aka athermal) effects, e.g. collective, coherent and resonant effects in bio–molecules and living cells a significant knowledge and data has been collected on this topic during the recent years. For the current state of the debate see for example [15, 22, 23]. At the same time, there is a growing number of experimental evidences demonstrating that a local microwave exposure stimulates tissue repair and regeneration, mitigates stress reactions and facilitates curing of various diseases, alleviates the effect of X-rays etc. [10, 20]. It is clear however that in each particular experimental situation the result of the irradiation of the bio-materials with sub–terahertz waves depends on a great number of physical factors such as: frequency; field strength; balance between incident, reflected and absorbed power; temporal dependence of the wave amplitude (e.g. constant or modulated, CW or intermittent pulsed wave), overall duration of the exposure and so on. The result depends strongly also on the physiological characteristics (tissue specific parameters depending on the treated organ, blood circulation), ambient environment, geometrical characteristics (e.g. plane or curved surface; volume to surface ratio) and many others. Experiments carried out in such multidimensional parameter space are difficult to compare and reproduce. Therefore, the only way to respond adequately to this problem is a systematic approach, which accounts for all these factors and experimental conditions.

It is expected that combining two different kinds of radiation therapy, namely the one based on irradiation by neutrons (more specifically BCNT) and the other which uses a collimated terahertz beam will lead not only to a *superposition effect* (i.e. an increased efficiency of the treatment due to the simultaneous or sequential application of two therapeutic techniques) but also to a *synergy effect*, i.e. effect, which is unique for the above-mentioned combination and could not be achieved if the techniques are applied separately. In order to prove this idea experimental equipment, which allows combining the irradiation by two beams (neutron and terahertz beam – a combination, which we call a *quantum beam* for short and following the definition given by the Japan Atomic Energy Agency Quantum Beam Directorate [24]) is necessary. This motivated us to initiate a research project directed towards the development of such hybrid dual−beam irradiation facility for an experimental cancer therapy. In this paper we present the conceptual design of such system and discuss the further steps of its feasibility study. It is believed that building this facility will allow to explore the above mentioned expected effects (superposition and synergy) of a novel combined therapy.

## 2. The gyrotron as a source of coherent terahertz waves for medical applications–current status and preceding studies

Gyrotrons are the most powerful sources of coherent electromagnetic radiation in the sub-millimeter wavelength region, operating in a CW (continuous wave) regime. In recent years, they have demonstrated a remarkable potential for further advancement toward higher (sub−terahertz and terahertz) frequencies and are being used in an ever increasing number of applications in the basic research and in the technologies [25–27]. In FIR FU, two families of devices namely, Gyrotron FU Series and Gyrotron FU CW Series covering a wide range from sub−THz to THz frequencies have been developed [26]. Their main characteristics and achievements are summarized in the Table 1 and Table 2. The tubes belonging to the FU Series are operated in a pulsed regime, while these from the FU CW Series generate a CW radiation. The most recently developed devices extend the FU CW Series to two novel branches, namely FU CW C and FU CW G, where the specification "C" stands for compact, and "G" for a gyrotron producing a Gaussian output beam in an internal mode converter. Some of these gyrotrons have operational characteristics and output parameters that are appropriate for experiments on novel medical technologies utilizing irradiation by microwaves and have already been used in some preceding similar studies.

The idea to use microwave (sub−Terahertz) radiation generated by a sub−millimeter wave gyrotron for treatment of cancerous cells has been pursued in a number of pioneering studies in the FIR FU Research Center [28–31]. During these investigations important issues (e.g. means for transmission of the radiation and its application to the treated area through a catheter antenna) have been worked out in detail. They have demonstrated clearly the following important advantageous features of the irradiation by sub-millimeter waves: (i) it is possible to localize the radiation to a small size spot on the living body tissue by the use of a waveguide vent antenna and Teflon rod antenna; (ii) increased absorption of the wave power because of the decreased reflections in the impedance-matching (anti−reflecting) dielectric layers at the used frequency range [32–34]; (iii) millimeter and sub−millimeter waves can be transmitted into the living body by a slim catheter; (iv) the dose of the deposited energy can be controlled easily by controlling precisely the output power of the gyrotron in a real time and measuring the reflected power from the tissue. Going to higher frequencies (terahertz beam instead of a millimeter wave beam), which is a combination of a hyperthermia treatment using gyrotron and a photodynamic treatment (PDT) with a photosensitizer and a laser for irradiation of an experimental tumor model [36].

Table 1. Gyrotron FU series

| Gyrotron | Frequency, THz | Characteristic features and achievements (The applications are distinguished by italic letters) |
|---|---|---|
| **FU I** | 0.038–0.220 | First high-frequency medium power gyrotron of the FU Series. Output power of 9 kW at 100 GHz. |
| **FU E** | 0.090–0.300 | *Radiation source for the first experiment on ESR* |
| **FU IA** | 0.038–0.215 | *Radiation source for plasma diagnostic at WT−3* |
| **FU II** | 0.070–0.402 | Studies on mode interactions (competition, cooperation etc.). *Radiation source for plasma scattering measurements at CHS in NIFS. Radiation source for XDMR experiments at ESRF.* |
| **FU III** | 0.100–0.636 | $3^{rd}$ harmonic single-mode operation. Amplitude modulation. Frequency step switching |
| **FU IV** | 0.160–0.847 | Frequency modulation. CW operation during long time periods with a high stability of the output power and frequency. Medical studies. |
| **FU IVA** | 0.160–0.889 | Highest frequency at third harmonic (a world record for a long period). *Radiation source for ESR experiments*. |
| **FU V** | 0.186–0.222 | CW operation for a long time using He−free superconducting magnet. High stability of the frequency and the amplitude. High mode purity |
| **FU VI** | 0.064–0.137 | Large−orbit gyrotron (LOG) with a permanent magnet. High harmonic operation up to $5^{th}$ harmonic of the cyclotron frequency. |

Table 2 Gyrotron FU CW series

| Gyrotron | Frequency range, THz | Power, W | Max. B, T | Characteristic features and achievements |
|---|---|---|---|---|
| **FU CWI** | 0.300 | 2300 | 12 | Advanced materials processing. Novel medical technologies. |
| **FU CW II** **FU CW IIA** | 0.110–0.440 | 20–200 | 8 | DNP−NMR at 600 MHz for protein research at Osaka University. Heating of Si substrate. Preliminary experiment on Bloch oscillation. |
| **FU CW III** | 0.130–1.080 | 10–220 | 20 | High−power THz technologies |
| **FU CW IV** | 0.131–0.139 | 5–60 | 10 | DNP-NMR at 200 MHz for analysis of polymer surface. |
| **FU CW V** | 0.2034 | 100–200 | 8 | Measurement of the hyper−fine splitting (HFS) of positronium. Radiation source for novel medical technologies. |
| **FU CW VI** | 0.393–0.396 | 50–100 | 15 | DNP−NMR at 600 MHz for protein research at Osaka University. |
| **FU CW VII** | 0.2037; 0.3953 | 200;50 CW | 9.2 | DNP−NMR at 300 and 600 MHz at Warwick University. |
| **FU CW VIIA** | 0.1315 0.395 | 200 | 8 | ESR echo experiment in the sub-THz region. |
| **FU CW VIII** | 0.100–0.350 | 100 | 8 | Pump and probe technique for XDMR at ESRF |

A typical experimental set–up of the preceding bio–medical experiments carried out in FIR FU is shown in Fig. 1. In this particular example the gyrotron FU IV was used as a radiation source for irradiation of a living tissue. It has a 12 T superconducting magnet and covers a wide frequency range from 199 GHz to 305 GHz, producing output powers of up to 20 W at the fundamental resonance of the cyclotron frequency. The radiation is transmitted through a cylindrical waveguide with an inner diameter of 28 mm and two bends and then to the catheter waveguide by a tapered waveguide section. The catheter waveguide is coupled to one or another antenna (waveguide vent antenna, WVA; Teflon rod dielectric antenna; conical horn antenna etc.). During the irradiation, the surface temperature is measured by an infrared radiation thermometer in order to estimate the energy absorbed in the tissue.

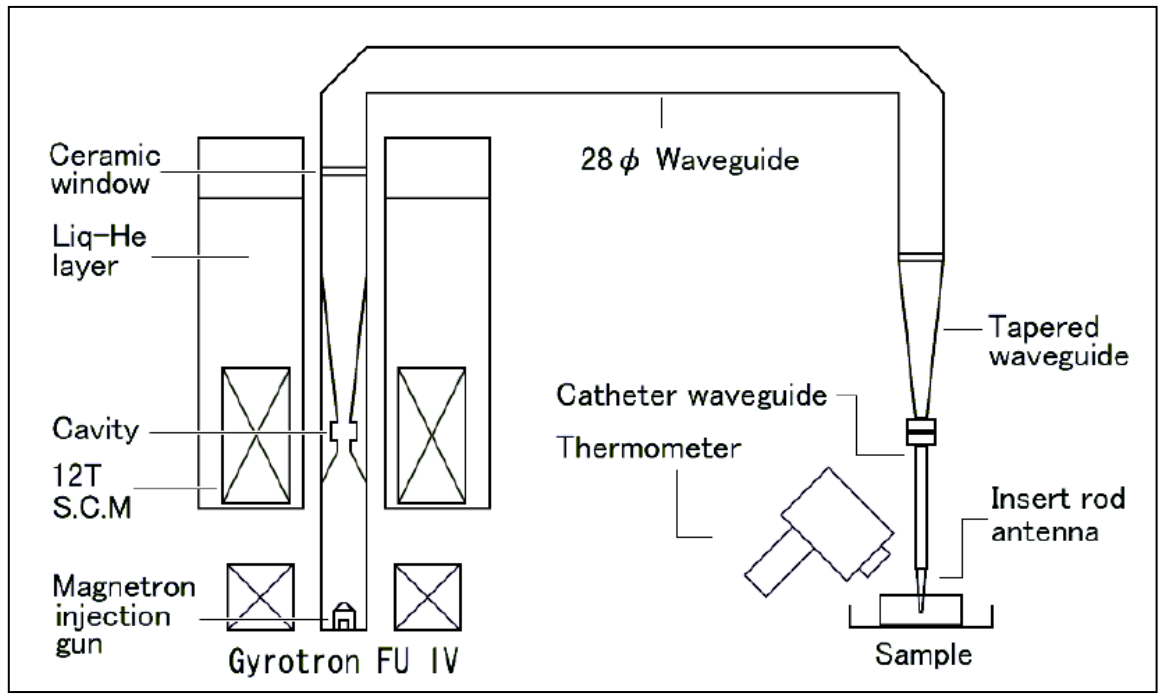

Fig.1. Experimental setup for millimeter wave irradiation of biological samples and invasion into living bodies using a gyrotron as a radiation source (adapted from [30]).

## 3. Conceptual design study of the hybrid dual–beam irradiation facility

A schematic block diagram and a functional scheme of the hybrid irradiation facility, which combines a neutron beam and a terahertz wave in a quantum beam, is shown in Fig. 2. Since the so-called dual beam is a result of the coalescence of two beams another appropriate term (alongside with a *quantum beam*) could be a *coalescent beam*.

The main components of the envisaged hybrid system are: (i) an accelerator and a neutron source; (ii) a gyrotron tube generating coherent terahertz radiation and a system for delivering the radiation to the treated zone of the patient's body consisting of a waveguide transmission system and a catheter antenna; (iii) systems for monitoring and control of the parameters of both the neutron and the terahertz beams; (iv) a diagnostic system for monitoring and control of the doses of irradiation; (v) a system for monitoring of the boron level in the blood of the patient as well as other bio–chemical parameters.

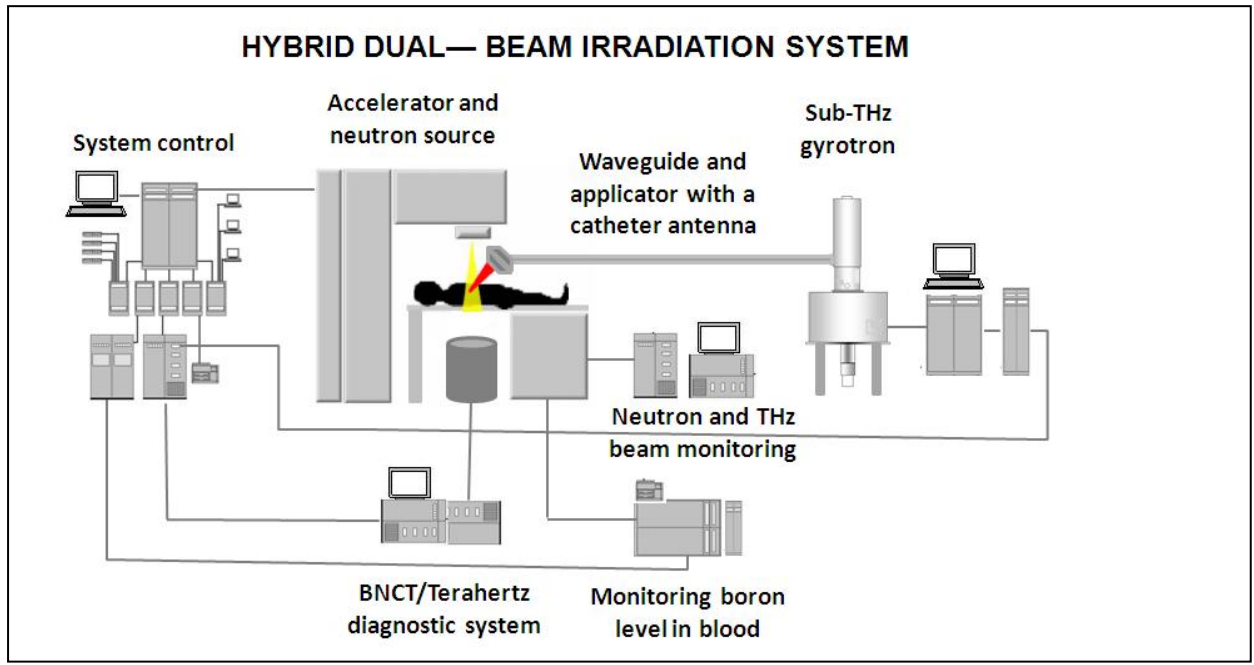

Fig. 2 Hybrid dual–beam irradiation facility: main components and a functional scheme

## 3.1. Simulation tools used for the conceptual computer aided design (CAD) of the main components

For a computer aided design (CAD) and an optimization of the neutron source and of the shielding (radiation protection) the code PHITS (**P**article and **H**eavy **I**on **T**ransport code **S**ystem) [37] is used. PHITS simulates the transport of neutrons, protons, heavy particles and photons. This advanced software package is also used for simulation of the energy distribution and doses to organs of voxel phantoms by the Monte Carlo method [38, 39]. As an example, Fig. 3 shows a 3D view and dose distribution in a voxel phantom calculated by PHITS. A treatment planning system for BNCT called JCDS is incorporated in PHITS as a Monte Carlo transport code. JCDS stands for **J**AEA (Japan Atomic Energy Agency) **C**omputational **D**osimetry **S**ystem [40, 41]. The scheme of the treatment planning using JCDS is illustrated in Fig. 4. It uses both CT (computer tomography) and MRI (magnetic resonance imaging) data for a precise evaluation of the absorbed radiation. Therefore, PHITS allows one to determine accurately the total doses received by the patient as a result of a combined modality therapy such as BNCT and X-ray therapy.

At FIR FU Research Center a problem-oriented software package GYROSIM (which stands for **Gyro**tron **Sim**ulation) has been developed for analysis, optimization and CAD of powerful high-frequency gyrotrons operating in a CW (continuous wave) mode and generating coherent radiation in the sub-terahertz and the terahertz frequency range. It is based on adequate and self-consistent physical models [42], implemented in highly portable and extensible numerical codes using efficient algorithms and programming techniques [43]. The structure of the package is shown in Fig. 5. Its components are specialized to the simulation of the main subsystems of the gyrotron tube, namely: (i) the electron−optical system (EOS), which forms a high-quality helical (or uni−axial, in the case of a large orbit gyrotron) electron beam with appropriate parameters; (ii) the electro−dynamical system (resonant cavity) in which the electron beam interacts with the high−frequency electromagnetic field and radiates a wave beam; and (iii) a quasi−optical system for converting the radiation to a Gaussian beam and its transmission to the target.

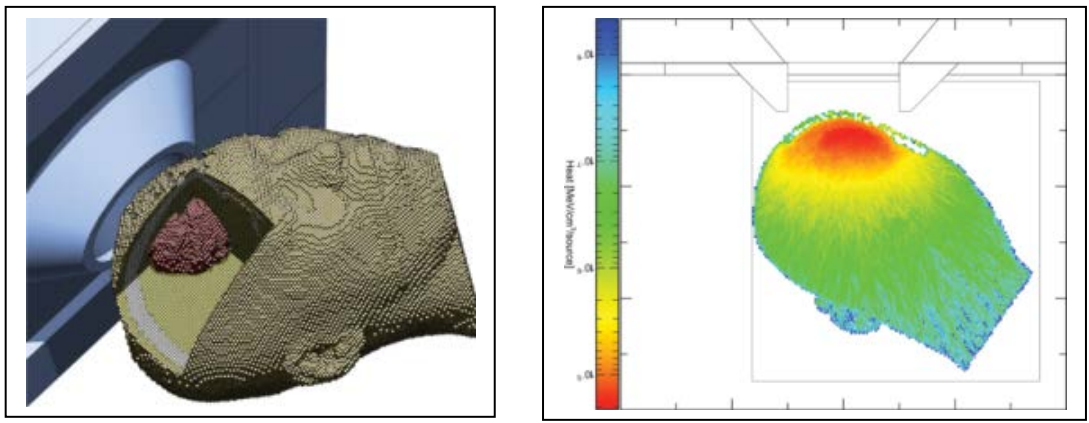

Fig. 3 3D view produced by PHITS (left), and dose distribution in the voxel phantom (right); available online at: http://phits.jaea.go.jp/AppsRaBNCT.html

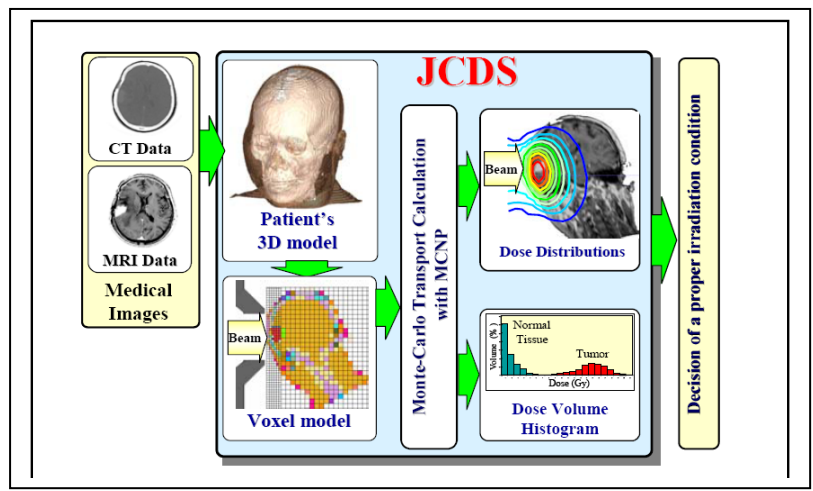

Fig. 4 Use of the JCDS for a treatment planning based on CT (computer tomography) and MRI (magnetic resonance imaging) data [40].

The process of the CAD of a gyrotron tube carried out using GYROSIM software package is shown in Fig.6. It includes several steps and optimization loops, repeated iteratively until a satisfactory configuration of the electrodes of the magnetron injection gun (MIG), magnetic field profile; shape and dimensions of the resonant cavity as well as the components of the quasi−optical system (beam launcher, mirrors of the mode converter etc.) are obtained.

Fig.7 shows some illustrative screen-shots from the numerical experiments carried out using GYROSIM package for analysis of different designs of gyrotrons which are under consideration as appropriate radiation sources for a dual− beam irradiation system.

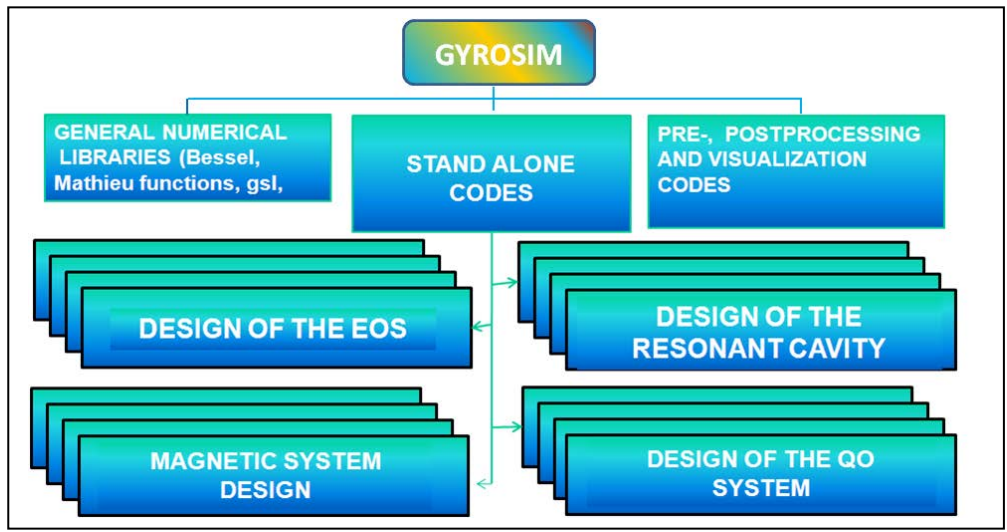

Fig. 5. Structure and content of the problem oriented software package GYROSIM for modeling and simulation of gyrotrons.

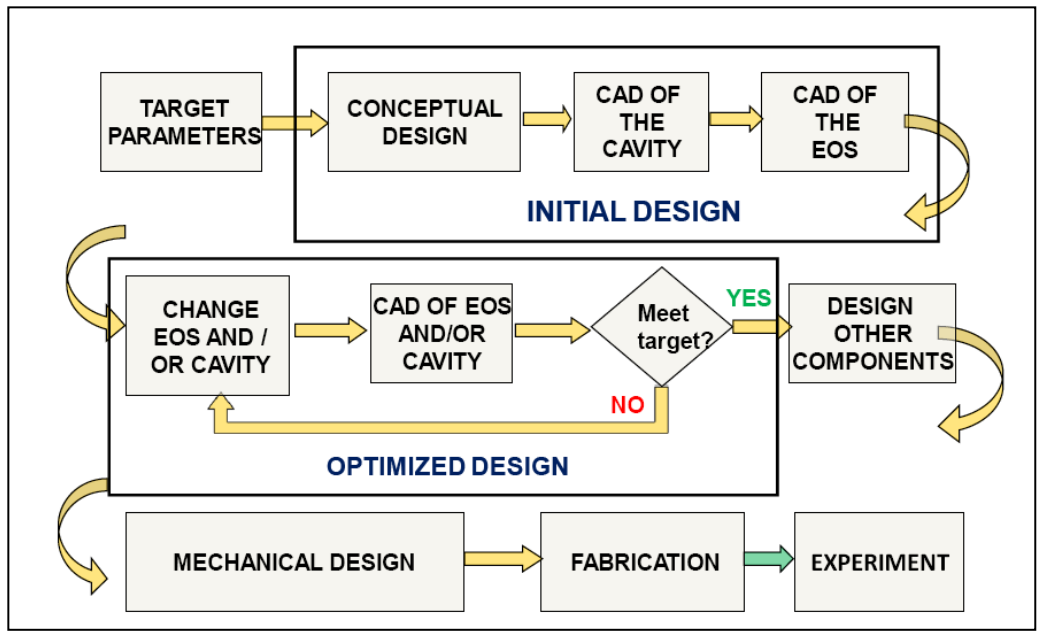

Fig.6 Flow chart of the computer−aided design using the components of the GYROSIM package.

## 3.2 Specification of the novel gyrotron and the first prototype

During the first stage of the feasibility and conceptual design study an appropriate area in the parameter space, where the prospective gyrotron for the dual−beam irradiation facility will operate has been selected. A specification of this radiation source is shown in Table 3. The main targeted parameters are the output frequency and the output power. The selected central frequency of 395 GHz corresponds to a wavelength of 0.76 mm, i.e. belongs to the sub−millimeter range. We however envisage an operation on a sequence of modes, and, as a result, frequency step tunability in a wide band around each frequency. This will allow us to use both millimeter and sub−millimeter waves in the conceived medical experiments with a quantum beam.

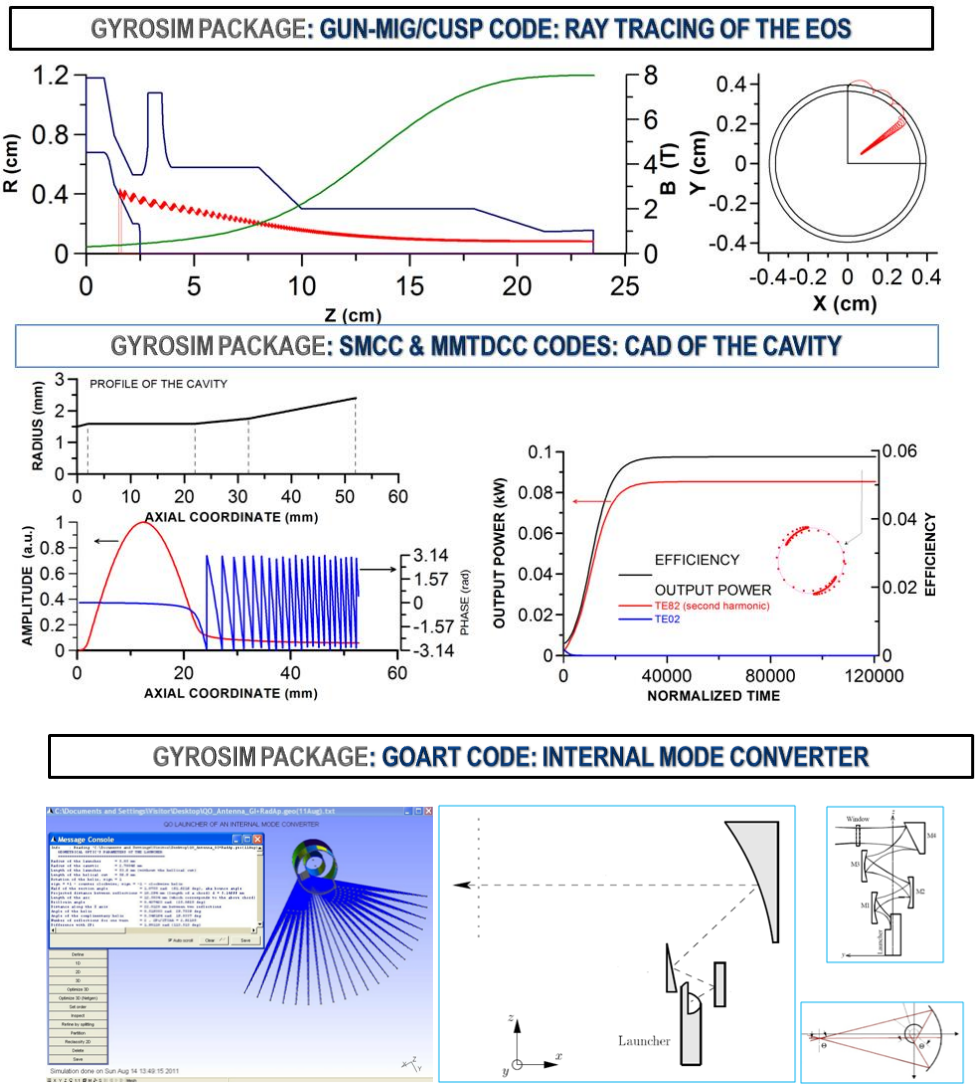

Figure 7 Illustrative examples from numerical experiments carried out using the software package GYROSIM: Trajectory analysis code GUN-MIG/CUSP (upper panel); single mode cavity code SMCC, and multimode time−dependent cavity code MMTDCC (middle panel); geometrical optics and analytical ray tracing code GOART (lower panel).

It should be noted that the output power levels are deliberately selected higher than those usually considered as necessary for a microwave irradiation in this spectral range. One reason for such a selection is that the transmission losses could vary significantly depending on the length of the system, and thus an adequate reserve of output power must be guaranteed. In any case, an attenuating section of the catheter waveguide is always present in the envisaged design and the power could be easily decreased if necessary. Having such relatively high output powers however, permits experimenting not only with low intensity irradiation but also with high intensity one. Recall that the power densities in the range 0.1–1 mW/cm$^2$ are regarded as non−thermal (i.e. not causing heating of the bio−medium of more than 0.1 K). The estimates cited in [9] show that a specific absorption rate (SAR) of 1 W/kg produces an increase of 1 °C in a human body, taking the thermoregulation into account. The data of other similar experiments summarized in [9] indicate also that the pulsed fields are able to produce a detectable change of the temperature at lower power levels than the continuous wave (CW).

Another consideration concerning the power level is related to the type of the application of the microwave energy. There are two varieties of the applicators, namely *invasive* and *noninvasive* ones. In the case of the noninvasive applicator (i.e. an applicator which does not penetrate the body and has not a contact with the tissue) the necessary power is generally higher due to the coupling loses. Since we plan to use both types in the experiments, the output power level in the specification (Table 3) is set according to the method which demands higher powers.

Table 1 Main requirements and specifications of the gyrotron as a radiation source for the dual−beam irradiation facility

| | | Description of the requirements | Appropriate range/solution |
|---|---|---|---|
| P A R A M E T E R S | Mandatory | <ul><li>Output power at the vacuum window</li><li>Power at the applicator</li><li>Central frequency</li><li>Low accelerating voltage $U_a$ and beam current $I_b$</li><li>Well collimated Gaussian beam</li><li>Effective coupling to the applicator</li><li>High stability of the output power</li><li>Stability of the frequency</li><li>Easy maintenance</li></ul> | 20-30 W<br>10-15 W<br>395 GHz (second harmonic)<br>$U_a$=12-15 kV;<br>$I_b$=150-250 mA<br>Internal mode converter<br><br>Transmission line with a corrugated waveguide<br>±1 mW<br>10 ppm<br>Cryo-free superconducting magnet ($B_{max}$=8 T) and computer control of the operational parameters |
| | Desirable | <ul><li>Frequency step tunability</li><li>Amplitude modulation</li></ul> | Control of the magnetic field in the cavity<br>Feedback controlled beam current, voltage, and magnetic field of the MIG |

For the concept under consideration, the possibility to irradiate the treated living tissue by an amplitude modulated sub–terahertz beam is an interesting opportunity. It is planned to study the difference (and possibly the advantages) of such a treatment as differ from the case of a constant wave. A technique for an amplitude modulation of the gyrotron output has been investigated in detail in a series of preceding experiments [44-45]. It is based on the changes of the pitch factor (velocity ratio) of the helical electron beam by controlling the anode voltage. Thus, a sub–system for both stabilization and modulation of the anode voltage based on a feed back control will be realized as a part of the overall control system of the tube.

As a whole, the development of this gyrotron follows our concepts (as described in [46]) for both innovative and standard designs of novel tubes using the available simulation tools for CAD outlined above. The selected tube type (from the nomenclature of the possible standard designs) is a sealed-off tube with a triode MIG and a regular cylindrical resonant cavity. An indispensible part of it is the internal mode converter which forms a well collimated Gaussian–like wave beam at the output window that is suitable for coupling with a low–loss corrugated waveguide of the transmission line. The tube will operate at both the fundamental and the second harmonic resonances of the cyclotron frequency on a number of modes (thus allowing a step tunability), including an excitation of a sequence of high–order axial modes for a continuous frequency tunability by controlling the magnetic field in the cavity.

The first prototype of such a tube has already been constructed at FIR FU. This gyrotron (FU CW CI) is one of newest members of the FU CW series and was conceived initially as a compact radiation source for DNP/NMR spectroscopy [26, 47]. Its operational characteristics are appropriate also for the planed dual-beam irradiation. Currently, the device is being tested, refurbished and prepared for commissioning to the neutron beam facility in the Takasaki site of JAEA. Its general view is shown in Fig. 8. FU CW CI is built using a cryogen-free superconducting magnet with maximum intensity of the magnetic field of 8 T. The total length of the tube (end-to-end) is only 1.02 m. Such compact and lightweight (almost table–top and portable) device could be easily embedded in the experimental irradiation facility. Initial experiments have shown that at a magnetic field of 7.2 T a single-mode operation on $TE_{26+}$ takes place at the second harmonic of the cyclotron frequency producing radiation with a frequency of 394.6 GHz. Several other modes have also been excited both at the fundamental and at the second harmonic operation.

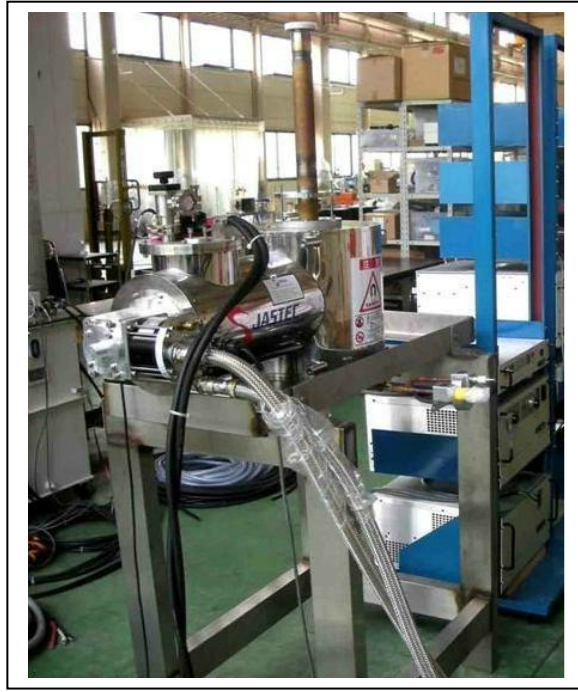

Fig. 8 Photo of the gyrotron FU CW C I (pre–prototype of the radiation source)

## 4. Further development of the necessary simulation tools for CAD of a hybrid dual–beam system for a radiation therapy

As already mentioned in the previous section, the available design tools (physical models and numerical codes) proved to be adequate for designing the two subsystems (for the neutron therapy and for the terahertz therapy) separately, following the principle of the decomposition. At the same time however, it should be noticed that the overall performance of the whole facility will depend strongly on the interplay between these two parts. That is why, the final design of both of them should proceed in parallel and in an iterative way till all target parameters are reached and all physical constrains are satisfied.

Bearing in mind that the two components of the quantum beam, namely the neutron beam and the terahertz beam are of different physical nature (corpuscular one and wave one, correspondingly) it is clear that such a task is not a trivial one. This applies specifically to the calculation of the total energy deposition in which both contributions (that of the neutron beam and that of the terahertz beam) have to be taken into account. Since the two beams are not interacting, they could, in principle, propagate through the space in overlapping regions without any interference. Due to their different nature however, the methods used for beam control and transport are also completely different. In the case of a terahertz beam produced by a gyrotron the transmission system and the beam propagation are usually analyzed from the point of view of the quasi-optical approach. In order to take into account the diffraction, which could influence strongly the radiation pattern of the antenna, either a full-wave treatment or a model based on the uniform geometrical theory of diffraction (UTD) (which is an extension of the geometrical theory of diffraction, GTD) must be used. At present, the development of such codes is in progress but the finalization of this work demands considerable resources (man power and time).

On the current stage of the project, which is a feasibility study preceding the final design of the whole system the following tasks are considered as essential ones for continuation of the development of the experimental facility: (i) Formulate a physical model for simulation of the radiation field and energy deposition in the targeted (irradiated) region produced as a result of the irradiation by a terahertz beam and implement this code as a module of the GYROSIM package. As a tentative working name for this module we will use the acronym TERAGYROBEAM; (ii) Conduct numerical experiments for testing and benchmarking of the developed module (TERAGYROBEAM) as well as for accumulation of data (data base) that characterize the correlation between the beam parameters and the distribution of both the reflected and the absorbed energy using different applicators and antennae; (iii) Interface TERAGYROBEAM with PHITS and develop a computational module (in PHITS) for calculation of the total energy distribution produced as a result of the simultaneous irradiation by the neutron and the terahertz beam as well as for evaluation of an equivalent (effective or biological) dose; (iv) Perform a series of numerical experiments using GYROSIM and PHITS for an optimized CAD of the components of the hybrid dual–beam system; (vi) Optimize the first prototype (FU CW CI) and manufacture a pilot version of a dedicated gyrotron for the hybrid irradiation system with an improved (turnkey) operational performance.

## 5. Conclusions and outlook

The outlined collaborative research project (and, in fact, interdisciplinary one) is focused on the realization of a unique and promising hybrid dual–beam irradiation system for advanced clinical studies on novel techniques for cancer treatment (malignant brain tumors, skin cancer etc.) utilizing a quantum beam. It has two main parts, namely a neutron source and a gyrotron producing a terahertz beam as well as a number of other auxiliary subsystems for monitoring and control of the irradiation process. The available simulation tools (physical models and numerical codes), for development of sub–terahertz and terahertz gyrotrons and neutron beam sources as well as for analysis of both the propagation and the interaction of the generated beams with the target have been used in this feasibility study in order to produce an initial conceptual design. Based on it, the first pre–prototype of the gyrotron has been developed and manufactured. Some tasks for further development of the codes that are necessary for an optimized final design of the whole irradiation facility have been formulated as well.

It should be mentioned that the terahertz beam produced by the gyrotron could be used not only for treatment of the irradiated tissue but also for microwave imaging and diagnostics [15, 48, 49]. In the modern therapeutic techniques the imaging (e.g. MRI, computer tomography etc.) is indispensible for detection and diagnostics, especially for the emerging image guided radiation therapy (IGRT) [50]. Therefore we plan to explore such possibilities and possibly integrate them in the next versions of the design.

It is believed that building such novel irradiation facility will open many new opportunities for experimenting with different therapeutic techniques utilizing a quantum beam and will facilitate the development of effective methods and tools for a cancer treatment.

## Acknowledgements

This work was supported partially by the Special Fund for Education and Research from Ministry of Education, Culture, Sports, Science and Technology (MEXT) in Japan and by SENTAN Project of Japan Science and Technology Agency (JST).

## References

[1] H. Hatanaka, Boron-Neutron Capture Therapy for Tumors, Springier (1986). ISBN 9780852006832

[2] Nakagawa Y, Pooh K, Kobayashi T, Kageji T, Uyama S, Matsumura A, Kumada H, Clinical review of the Japanese experience with boron neutron capture therapy and a proposed strategy using epithermal neutron beams, *J Neuro-oncol*, **62** (2003) 87–99.

[3] Buchholz TA, Laramore GE, Wootton P, Livesey JC, Wilbur DS, Risler R, Phillips M, Jacky J, Griffin TW., Enhancement of fast neutron beams with boron neutron capture therapy. A mechanism for achieving a selective, concomitant tumor boost, *Acta Oncol.*, **33** (1994) 307-313.

[4] Barth, RF, Grecula, JC, Yang, W, Rotaru, JH, Nawrocky, M, Gupta, N, Albertson, BJ, Ferketich, AK, Moeschberger, ML, Coderre, J, and Rofstad, EK., Combination of boron neutron capture therapy and external beam radiotherapy for brain tumors, *Int. J. Radiat. Oncol. Biol. & Physics*, **58** (2004) 267-277.

[5] R. Barth, Boron neutron capture therapy at the crossroads: Challenges and opportunities, *Applied Radiation and Isotopes*, **67** (2009) 3-6.

[6] D. W. Nigg, Neutron Sources and Applications in Radiotherapy – A Brief History and Current Trends (12th International Symposium on Neutron Capture Therapy, 8-12 Oct, 2006, Takamatsu, Japan) INL/CON-06-11615 PREPRINT.
 (Available online at: http://www.inl.gov/technicalpublications/Documents/3395012.pdf)

[7] Compact neutron generators. http://www.lbl.gov/tt/techs/lbnl1764.html#minitu

[8] K.-N. Leung, T. Lou, J. Puireijonen, Neutron tubes, US Patent US7342988 (11 Mar 2008).

[9] A. Rosen, M. A. Stuchly, A.V. Vorst, Applications of RF/Microwaves in Medicine, *IEEE Trans. On Microwave Theory and Techniques*, **50** (2002) 963-974.

[10] S. Banik, S. Bandyopadhyay, S. Ganguly, Bioeffects of microwave–a brief review, *Bioresource Technology*, **87** (2003) 155-159.

[11] N.D. Devyatkov, M.B. Golant, O.V. Betskii, Millimeter Waves and their Role in the Process of Vital Activity, 1991 (Moscow: Radio I Svyaz') (in Russian).

[12] S.A. Reshetnyak, V.A. Shcheglov, V.I. Blagodatskikh, P.P. Gariaev, M.Yu. Maslov, Mechanisms of Interaction of Electromagnetic Radiation with a Byosystem, *Laser Physics*, **6** (1996) 621-653.

[13] C. Furse, D. A. Christensen, C. H. Durney, Basic introduction to bioelectromagnetics, Second ed., CRC Press (2009).

[14] M. C. Ziskin, Physiological Mechanisms Mechanisms Underlying Millimeter Wave Therapy, In: Bioelectromagnetics, current concepts: the mechanisms of the biological effect of extremely high power pulses, S.N. Ayrapetyan and M.S. Markov (eds.), (Springer, 2006) 241-251.

[15] P.H. Siegel, Terahertz Technology in Biology and Medicine, *IEEE Trans. Microwave Theory and Techniques*, **52** (2004) 2438-2447.

[16] A. R. Orlando, G. P. Gallerano, Terahertz Radiation Effects and Biological Applications, *J. Infrared Millimeter and Terahertz Waves*, **30** (2009) 1308-1318.

[17] G. J. Wilmink, J. E. Grundt, Invited Revew Article: Current State of Research on Biological Effects of Terahertz Radiation, *J. Infrared Millimeter and Terahertz Waves*, **32** (2011) 1047-1122.

[18] H. Frohlich, The extraordinary dielectric properties of biological materials and the action of enzymes, *Proc. Nat. Acad. Sci.*, **72** (1975) 4211–4215.

[19] G. J. Hyland, Physical basis of adverse and therapeutic effects of low intensity microwave radiation, *Indian Journal of Experimental Biology*, **46** (2008) 403–419.

[20] A. G. Pakhomov, Current state and implications of research on biological effects on millimeter waves: a review of the literature, *Bioelectromagnetics*, **19** (1998) 393-413.

[21] K. Mileva, B. Georgieva, N. Radicheva, About the biological effects of high and extremely high frequency electromagnetic fields, *Acta Physiol Pharmacol Bulg.*, **27** (2003) 89-100. Review. PubMed PMID: 14570154.

[22] D.T. Pooley, Bioelectromagnetics, complex behavior and psychotherapeutic potential, *Q J Med.*, **103** (2010) 545–554.

[23] J.R. Reimers, L.K. McKemmish, R. H. McKenzie, A.E. Mark, N.S. Hush, Weak, strong, and coherent regimes of Frohlich condensation and their application to terahertz medicine and quantum consciousness, *Proc. Nat. Acad. Sci. of USA*, **106** (2009) 4219-4224.

[24] Japan Atomic Energy Agency Quantum Beam Science Directorate:
http://qubs.jaea.go.jp/en_index.html

[25] T. Idehara, T. Saito, I. Ogawa, S. Mitsudo, Y. Tatematsu, S. Sabchevski, The potential of the gyrotrons for development of the sub-terahertz and the terahertz frequency range — A review of novel and prospective applications.- *Thin Solid Films*, **517** (2008) 1503-1506.

[26] Sabchevski S., Idehara T., Development and Applications of High—Frequency Gyrotrons in FIR FU Covering the sub-THz to THz Range, *Journal of Infrared, Millimeter, and Terahertz Waves,* Published on-line first on 7 Jan 2012, 28 pages.

[27] M. Thumm, State-of-the-Art of High Power Gyro-Devices and Free Electron Masers, Update 2011, KIT SCIENTIFIC REPORTS **7575**, (2011) 128 pages.
ISSN 1869-9669

[28] T. Tatsukawa, A. Doi, M. Teranaka, H. Takashima, F. Goda, T. Idehara, I. Ogawa, S. Mitsudo, T. Kanemaki, Development of Submillimeter Wave Catheter Transmitting a Gyrotron Output for Irradiation on Living Bodies, *Int. J. Infrared and Millimeter Waves,* **21** (2000) 1155-1167.

[29] T. Tatsukawa, A. Doi, M. Teranaka, H. Takashima, F. Goda, T. Idehara, I. Ogawa, T. Kanemaki, S. Nishizawa, Submillimeter Wave Irradiation of Living Bodies using a Gyrotron and a Catheter, *Jpn. J. Appl. Phys*., **41** (2002) 5486–5489.

[30] T. Tatsukawa, A. Doi, M. Teranaka, T. Idehara, T. Kanemaki, I. Ogawa, S.P. Sabchevski, Submillimeter Wave Irradiation on Living Bodies Using Catheter Waveguide Vent Antennae with Dielectric Rod and Sheet. In: NANOscale Magnetic Oxides and Bio-World, Edited by I. Nedkov and Ph. Tailhades (Heron Press Ltd., Sofia) pp. 123-138 (2004).

[31] T. Tatsukawa, A. Doi, M. Teranaka, H. Takashima, F. Goda, S. Watanabe, S. Mitsudo, T. Idehara, T. Kanemaki, T. Namba, Millimeter Wave Irradiation and Invasion into Living Bodies Using a Gyrotron as a Radiation Source, Proc. Int. Workshop on Strong Microwaves in Plasma, **2** (2006) 727-731.

[32] T. Tatsukawa, A. Doi, M. Teranaka, H. Takashima, F. Goda, S. Watanabe, T. Idehara, T. Kanemaki, T. Namba, Microwave invasion through anti-reflecting layers of dielectrics at millimeter wave irradiation to living bodies, *Int. J. Infrared and Millimeter Waves,* **26** (2005) 591-606.

[33] M. Teranaka, A. Doi, T. Tatsukawa, S. Mitsudo, T. Saito, T. Idehara, T. Kanemaki, T. Namba, Millimeter wave irradiation and invasion into living bodies by the anti-reflecting effect, Proc. 32$^{nd}$ Int. Conference Infrared, Millimeter and Terahertz Waves IRMMW-THz 2007 (2-9 Sept. 2007, Cardiff, UK) 571-572.

[34] M. Teranaka, A.Doi, I. Ogawa, T. Saito, T. Idehara, T. Tatsukawa, Millimeter wave irradiation and invasion into living bodies using AR waveguide vent antennas and Gyrotron, Proc. 33$^{rd}$ Int. Conference Infrared, Millimeter and Terahertz Waves IRMMW-THz 2008 (15-19 Sept. 2008, Pasadena, CA) 1-2.

[35] N. Miyoshi, Y. Fukunaga, I. Ogawa, T. Idehara, Application for hyperthermia treatment of an experimental tumor using a gyrotron (107, 203 GHz), Proc. 34$^{th}$ Int. Conference Infrared, Millimeter and Terahertz Waves IRMMW-THz 2009 (21-25 Sept. 2009, Busan, Korea), 1-2.

[36] N. Miyoshi, S. Ito, I. Ogawa, T. Idehara, Combination treatment of hyperthermia and photodynamic for experimental tumor model using gyrotron (107, 203 GHz), Proc. 35$^{th}$ Int. Conference Infrared, Millimeter and Terahertz Waves IRMMW-THz 2010 (5-10 Sept. 2010, Rome, Italy) 1-2.

[37] K. Niita, N. Matsuda, Y. Iwamoto, H. Iwase, T. Sato, H. Nakashima, Y. Sakamoto and L. Sihver, "PHITS: Particle and Heavy Ion Transport code System, Version 2.23", JAEA-Data/Code 2010-022 (2010).

[38] D. Satoh, K. Sato, F. Takahashi, A. Endo, Monte Carlo simulation using Japanese voxel phantoms to analyze the contribution of particle types and their energy distributions to organ doses upon external neutron exposure, *J. Nucl. Sci. Thechnol*., **47** (2010) 62-69.

[39] T. Sato, Y. Kase, R. Watanabe, K. Niita and L. Sihver, Biological dose estimation for heavy ion therapy using an improved PHITS code coupled with the micro-dosimetric kinetic model, *Radiat. Res*., **171**, (2009) 107-117.

[40] H. Kamada, K. Yamamoto, A. Matsumura, T. Yamamoto, Y. Nakagawa, Development of JCDS, a computational dosimetry system at JAEA for boron neutron capture therapy, *J. Physics: Conference Series*, **74** (2007) 0122010.

[41] H. Kamada, T. Nakamura, M. Komeda, A. Matsumura, "Development of a multi-modal Monte-Carlo radiation treatment planning system combined with PHITS," AIP Conf. Proc. 1153 377-387 (2009).

[42] S. Sabchevski, T. Idehara, M. Glyavin, I. Ogawa, S. Mitsudo, Modelling and simulation of gyrotrons, *Vacuum*, **77** (2005) 519-525.

[43] S. Sabchevski, T. Idehara, T. Saito, I. Ogawa, S. Mitsudo, Y. Tatematsu, Physical Models and Computer Codes of the GYROSIM (GYROtron SIMulation) Software Package, FIR Center Report FIR FU-99 (Jan. 2010). Available on-line at: http://fir.u-fukui.ac.jp/FIR_FU99S.pdf

[44] T. Idehara, Y. Shimizu, S. Makino, K. Ichikawa, T. Tatsukawa, I. Ogawa and G.F. Brand, High-frequency, amplitude modulation of a submillimeter wave, medium power gyrotron *Phys. Plasmas,* **1** (1994) 461-463.

[45] T. Idehara, Y. Shimizu, S. Makino, K. Ichikawa, T. Tatsukawa, I. Ogawa and G.F. Brand, Amplitude modulation of submillimeter wave gyrotron output, *Int. J. Infrared and Millimeter Waves*, **18** (1997) 391-403.

[46] S. Sabchevski, T. Idehara, Development of Sub-Terahertz Gyrotrons for Novel Applications, Proc. 36 Int. Conf. on Infrared, Millimeter and Terahertz Waves IRMMW-THz 2011 (2-7 Oct, 2011, Houston, USA) 1-2 (Th5.15).

[47] J. C. Mudiganti, T. Idehara, Y. Tatematsu, R. Ikeda, Y. Yamaguchi, T. Saito, I. Ogawa, S. Mitsudo, T. Fujiwara, Y. Matsuki, K. Ueda, Design of a 394.6 GHz Compact Gyrotron FU CW CI for 600 MHz DNP-NMR Spectroscopy, Proc. 36 Int. Conf. on Infrared, Millimeter and Terahertz Waves IRMMW-THz 2011 (2-7 Oct, 2011, Houston, USA) 1-2 (Tu5.8.1).

[48] Z. D. Taylor*,* R. S. Singh*,* D. B. Bennett*,*P. Tewari, C. P. Kealey, N. Bajwa, M. O. Culjat*,* A. Stojadinovic, H. Lee*,* J.-P. Hubschman, E. R. Brown*,* W. S. Grundfest*,* THz Medical Imaging: *in vivo* Hydration Sensing, *IEEE Trans. Terahertz Sci. and Technology*, **1** (2011) 201-219.

[49] Y. Sun, M. Y. Sy, Yi-X. J. Wang, A. T. Ahuja, Y.-T. Zhang, E. Pickwell-MacPherson, A promising diagnostic method: Terahertz pulsed imaging and spectroscopy, *World J Radiol.,* **3** (2011) 55-65.

[50] A.J. Mundt, J.C. Roeske, Image-Guided Radiation Therapy: A Clinical Perspective (PHPH-USA Ltd., 2011) 650 pages. ISBN 978-1-60795-042-4.